\documentclass[preprint,12pt,authoryear]{elsarticle}

\usepackage[final]{changes}

\usepackage{amssymb}
\journal{Icarus}

\begin{document}
%\linenumbers
\begin{frontmatter}



\title{Characterization of the M1 and M2 layers in the undisturbed Martian ionosphere \replaced{through solar minimum leading to solar maximum}{at solar minimum} with MAVEN ROSE}


\author[inst1]{Jennifer Segale}

\affiliation[inst1]{organization={Astronomy Department, Boston University},%Department and Organization
            addressline={725 Commonwealth Avenue}, 
            city={Boston},
            postcode={02215}, 
            state={MA},
            country={USA}}

\author[inst2]{Marianna Felici}

\author[inst1,inst2]{Paul Withers}

\affiliation[inst2]{organization={Center for Space Physics},%Department and Organization
            addressline={725 Commonwealth Avenue}, 
            city={Boston},
            postcode={02215}, 
            state={MA},
            country={USA}}

\author[inst3]{Shannon Curry}

\affiliation[inst3]{organization={Laboratory for Atmospheric and Space Physics, University of Colarado Boulder},%Department and Organization
            addressline={1234 Innovation Drive}, 
            city={Boulder},
            postcode={80303}, 
            state={CO},
            country={USA}}

\begin{abstract}
%What is the main overarching issue that the study described in the paper will help resolve in the future?Which critical hurdle towards the main overarching issue has the study described in the paper cleared?
%What is new in the study that enabled it to clear the hurdle? What new is learned in the study? What is the implication of the new results in the context of future 
%endeavors towards the main overarching issue?

\deleted{Is the Martian ionosphere symmetric in local time? Does the ionosphere change between aphelion, when there is no dust in the atmosphere, compared to perihelion, when the temperature rises and there is dust in the air even if no storm is blowing?}
  We utilise data from the MAVEN Radio Occultation Science Experiment \citep{Withers:2020um} - with unprecedented coverage in solar zenith angle - \deleted{to answer these questions and} to isolate the effects that local time and season induce on the \added{photochemical} ionosphere \replaced{of Mars around}{at} solar minimum\added{, leading to solar maximum}.
219 out of the  1228 electron density profiles of the Martian \added{undisturbed - by solar events or dust storm - dayside} ionosphere collected by MAVEN ROSE between July 2016 and December 2022  show\deleted{, besides the ever-present dayside M2 layer} a distinct M1 layer \deleted{right} below \added{the M2 layer}. This allowed us to study the behavior of \added{both} the M2 and M1 peak \replaced{densities and altitudes}{density and altitude} as a function of solar zenith angle, and, \deleted{also,} for the first time, \replaced{to be able to separate these trends in dusk and dawn local time, as well as by Southern Spring and summer versus Southern Fall and Winter}{local time, and Martian season}. We find that the M1 layer at \replaced{small}{low} SZA can occur at altitudes lower than 100 km; that the peak altitudes and densities of both the M2 and M1 layers \added{at dawn} change more with season than \replaced{they do at dusk}{at the dusk ionosphere}; and that the M2 peak density decreases at a faster rate than the M1 with SZA. \deleted{This study provides a baseline to accurately characterise the photoproduced Martian ionosphere at solar minimum.} % I want to change this sentence but not sure exactly what to say - results give us an idea of what we can expect the peak altitudes and densities of ionospheric layers to be at different times and seasons
\end{abstract}

\begin{highlights}
\item  \replaced{There is more seasonal variability in the dawn ionosphere than the dusk ionosphere;}{M1 and M2 peak densities and altitudes at the dawn ionosphere change more with season than the ones at dusk;}
\item The M1 layer is present at small SZA and, when present, is located $ \sim20$ km below the M2 layer;
\item The ratio of M1 over M2 peak densities increases \replaced{with}{in} SZA, and the two layers get farther from one another.
\end{highlights}


\end{frontmatter}


\section{Introduction}
\label{sec:intro}
%appendix~\ref{sec:sample:appendix}.
Historically, radio occultation observations have been a well utilised method to study planetary ionospheres, especially the ionosphere of Mars, because of their broad vertical coverage, good vertical resolution, and low density uncertainty \citep[e.g.,][and references therein]{Withers:2020um}. Electron density profiles of the Martian dayside ionosphere generally show two peaks, with one of lesser density just below the other. Each peak indicates a different layer. The upper layer is referred to as the M2 layer, while the lower layer is called the M1 layer. 

The M2 layer is where the largest electron density peak occurs and is usually located at an altitude between about 120 and 180 km \citep[e.g.,][]{bougher2017}. \replaced{The M2 layer}{It} is produced by ionization due to extreme ultraviolet (EUV) photons. In general, as solar zenith angle (SZA) increases, the peak electron density value in the M2 layer decreases, while the altitude of the peak increases \citep[e.g.,][and reference therein]{fallows2015a}. This trend follows closely the one expected from an idealized photochemical theory  \citep{chapman1931a, chapman1931b}. %The M2 layer is also affected by many other factors, including solar irradiance and magnetic fields. It has been found that as solar activity increases, peak electron density also increases. Additionally, in regions with strong crustal magnetic fields, electron densities have been found to be higher \citep{Withers:2009aa}.
The M1 layer is situated below the M2 layer. \replaced{The M1 layer}{It} appears on electron density profiles as a small peak or shoulder below the M2 peak, but \deleted{it} is not always present or visible, and its behavior is less understood [Withers 2023, SUBMITTED]. \replaced{The M1 layer}{It} forms by ionization from soft X-rays \added{and Auger electrons}, as opposed to EUV photons for the M2 layer \citep[e.g.,][]{bougher2017}. Similar to the M2 layer, the peak electron density of the M1 layer decreases as the SZA increases. The altitude of the peak increases as SZA increases, as it does for the M2 layer, but previous studies\added{, conducted during solar maximum,} show \replaced{this change}{it} occurring at a slower rate \added{than it does for the M2 layer} \citep[][and references therein]{fallows2015a}. %The M1 layer can be difficult to identify as it varies greatly in altitude and density and its presence can be very uncertain. %However, there is evidence that electron densities around the M1 layer increase significantly during solar flare activity \citep[e.g.,][]{Withers:2009aa}. %In general, the peak density of the M2 and M1 layers are greater during high solar activity \citep{}.

The peak altitudes and densities of \replaced{the M1 and M2 layers}{these layers} change depending on various factors\replaced{, such as}{like} phase in the solar cycle \citep[e.g.,][]{Withers:2023aa}\replaced{: while the scale height above the M2 peak in the photochemical ionosphere is quite similar going from solar maximum to solar minimum, the M1 and M2 peak densities increase from solar minimum to solar maximum, because of the increased solar irradiance. Other factors that can affect the M1 and M2 layers, and the ionosphere in general, include }{,  }solar activity \citep[e.g.,][]{Hantsch:1990} [Felici, COMPANION MANUSCRIPT], solar flares \citep[e.g.,][]{mendillo2006}, dust storms  \citep[][]{Kliore:1972, felici2020}[Felici, COMPANION MANUSCRIPT], and local time sector \citep[e.g.,][]{Pilinski:2019aa, Felici:2022aa}. The changes that these factors induce on the peak densities and altitudes are better characterized for the M2 layer, and less studied for the M1 layer \citep[e.g.,][and references therein]{fox-yeager2006, fox-yeager2009, fallows2015a} [Withers 2023, SUBMITTED].  \added{The nightside ionosphere, beyond the terminator, will not be considered in this study, aside from helping determine the location of the terminator: photoionization is absent on the deep nightside, and instead horizontal transport of dayside photo-produced plasma and ionisation due to precipitating particles are the origin of electron densities larger than zero that can be detected beyond the terminator \citep[e.g.,][]{withers2012b, Nemec:2010}. }

%The features of the M1 layer can vary so much that sometimes it is not detected. 

\deleted{The Radio Occultation Science Experiment (ROSE)  was added to the MAVEN  science payload in 2016. Since then, ROSE has collected more than 1000 electron density profiles, and a good number of these show both M1 and M2 layers. While geometric constraints limit radio occultation observations at Mars to Solar Zenith Angles (SZA) between $\sim$ 45$^{\circ}$ and $\sim$136$^{\circ}$, making it impossible for this technique to observe subsolar peak altitudes, ROSE has been able to cover that range nearly completely down to $\sim$47$^{\circ}$ [Withers et al. 2023, SUBMITTED], in addition to a comprehensive coverage in latitude and longitude. } %This allowed us to make comparisons between the M1 and M2 layers under a broad range of conditions and investigate how their peak altitude and density changes, independently and relative to one another.

Previous studies that used \added{dayside} radio occultation data\replaced{ -}{,} very few of which focused \deleted{also} on the M1 layer\replaced{ -}{,} \deleted{besides the M2,} have found some variation in both M1 and M2 peak densities and altitudes at similar SZAs \citep[e.g.,][]{yao2019, fallows2015a} \added{consistent with what is expected in an idealised photochemical theory \citep{chapman1931a,chapman1931b}}, and their extrapolation of subsolar peak altitude and density led to different results \citep[e.g.,][]{nielsen2006, morgan2008, fox-yeager2009, fox2012}\added{, some of which we report in Tables \ref{table:2} and \ref{table:3} to contextualise our results in Section \ref{sec:results}}; however, \added{while longitudinal and thermal tide dependence \citep[][]{bougher2004, fox-weber2012}, and irradiance dependence \citep[e.g.][]{fox-yeager2009} were previously considered,} the M1 and M2 peak \replaced{density and altitude changes with SZA}{densities and altitudes} were never\replaced{, to our knowledge, }{before} separated in local time sector or season to quantify their dependence \replaced{on}{from} these factors\added{, which we expect have an effect \citep[e.g.,][]{Pilinski:2019aa, Felici:2022aa} on the idealised photochemical ionosphere \citep{chapman1931a, chapman1931b, Fallows:2015sza} (please see Section \ref{sec:methods}). In other words, we expect that at fixed solar zenith angles, the ionosphere at southern spring and summer will be different from the ionosphere at southern autumn and winter, and the ionosphere at dawn will be different than the ionosphere at dusk. We utilise M1 and M2 peak densities and altitudes as proxies to quantify "how different" the ionosphere is in such circumstances. Additionally, quantitatively characterising the changes that local time and season induce on the Martian ionosphere when undisturbed - namely no solar events or dust storms are affecting the ionosphere - is paramount to understanding and quantifying ionospheric variability in the presence of external triggers and during a more active phase of the solar cycle.}%, leading to even questioning the existence of the M1 layer at low SZAs \citep{mayyasi2015}. %However, many studies focus on the M2 layer over the M1. There has been only two prior examinations of the M1 layer at SZAs less than 70 degrees \citep{yao2019} and [XXX Withers et al. 2023, SUBMITTED], also using ROSE data. Since this study was published there have been more occultations at lower SZAs. Other studies have also looked into the effects of factors such as local time \citep{mayyasi2015}, season \citep{morgan2008}, and solar activity \citep{fox-yeager2009, fox2012}.

The thorough SZA coverage of \replaced{the Radio Occultation Science Experiment (ROSE) \citep{Withers:2020um}}{ROSE} allowed us to make comparisons between the M1 and M2 layers under different \added{solar longitude (L$_{S}$) and local solar time (LST)} conditions and investigate how \replaced{the M1 and M2}{their} peak altitude and density change\deleted{s}, independently and relative to one another. 

\added{The Radio Occultation Science Experiment (ROSE) \citep{Withers:2020um} was added to the MAVEN \citep{Jakosky:2015} science payload in 2016. Since then, ROSE has collected more than 1000 electron density profiles, and a good number of these show both M1 and M2 layers. 
While geometric constraints limit radio occultation observations
at Mars to Solar Zenith Angles (SZA) between $\sim$ 45$^{\circ}$ and $\sim$136$^{\circ}$, making it impossible for this technique to observe subsolar peak altitudes \citep[e.g.][]{hinson1999, Tamburo:2023aa}, ROSE has been able to cover that range nearly completely down to $\sim$47$^{\circ}$ [Withers et al. 2023, SUBMITTED], in addition to \replaced{a}{an} comprehensive coverage in latitude and longitude.}

\replaced{The}{In this paper, we} focus \added{of this study is} on the effects that Local Solar Time (LST) - dawn versus dusk local time sectors - and season - dust season\replaced{ near}{,} perihelion\deleted{,} versus non-dust-season\replaced{ near}{,} aphelion - have on the trends that peak altitude and density are expected to have with SZA from the idealized photochemical theory. \added{To better isolate these effects, we focus on the "undisturbed" ionosphere: when there are no known events external to the ionosphere (e.g. ICMEs, dust storms) causing perturbations.} \replaced{We}{In fact, we} expect the dusk ionosphere, exposed for longer to light, to be different, and extend to higher SZA angles, than the dawn ionosphere, that it is just coming out from night. We also expect dust season\deleted{, on itself,} to have effects on the ionosphere: at perihelion, in fact, some amount of dust is lofted in the atmosphere, regardless of whether or not there are dust storms \added{\citep{Montabone:2020aa}}. 

\added{Because of the effects that increased solar irradiance can have on ionospheric conditions, we feel it is important to note the phases of the solar cycle in which data was obtained. The ROSE electron density profiles utilised in this study  were collected during the less active half of the solar cycle, from just before solar minimum to just before solar activity ramps up approaching solar maximum (see Figure \ref{fig_4} for a visualization of when data was acquired). } %\added{This work is also most accurate for times of low solar activity, as the measurements used were taken during the less active half of the solar cycle.}

\deleted{Therefore, the aim of this study is to offer a deeper characterization of the behavior of both the M1 and M2 layers, studying not only M1 and M2 peak densities and altitudes as a function of SZA, utilising the complete coverage that ROSE offers, but also how those trends change with season and local time sector, to fully investigate causes for the aforementioned variation. }%The dataset used in this study was collected by  the MAVEN spacecraft’s Radio Occultation Science Experiment (ROSE) \citep{Withers:2020um}. 
This paper is organised as follows: in Section \ref{sec:methods} we will illustrate how the data was filtered and analyzed; in Section \ref{sec:results} we will report and discuss the results of this study; in Section \ref{sec:conclusions} we summarize and report our conclusions.

\section{Methods}
\label{sec:methods}
We began by manually examining  1228 electron density profiles collected by MAVEN ROSE \citep{Withers:2020um} between July 2016 and December 2022. The data covered a SZA range between $\sim$47$^{\circ}$ and $\sim$136$^{\circ}$, latitudes from  -85.4$^{\circ}$ to 88.9$^{\circ}$, \added{and had} comprehensive coverage in longitudes and L$_{S}$. 

\added{We applied four criteria to select the profiles which would be included in this study. First, even though ROSE profiles collected in 2023 were available on the PDS when this manuscript was written, we limited this study to the lower part of the solar cycle, only approaching the higher solar activity of solar maximum. This choice allowed to include enough profiles distributed in the LST and L$_{S}$ sectors we wanted to study, while avoiding high solar activity that caused noticeable outliers in the data.} 

\replaced{Second, we}{We} chose to include in this study only profiles that we deemed "complete" in altitude, namely profiles \replaced{with full altitude coverage}{that, at the very least, covered altitudes} between 75 km and 250 km, \added{in order }to have a complete view of predicted altitudes for the M1 and M2 layers\replaced{. Additionally, coverage of altitudes above the main ionospheric layers allows for better baseline corrections during the process of generating electron density profiles, making these profiles more reliable for determining M2 and M1 peak altitudes and densities}{, and include robust baselines} \citep{Withers:2020um}.

\begin{figure}
	\centering 
	\includegraphics[width=1\textwidth]{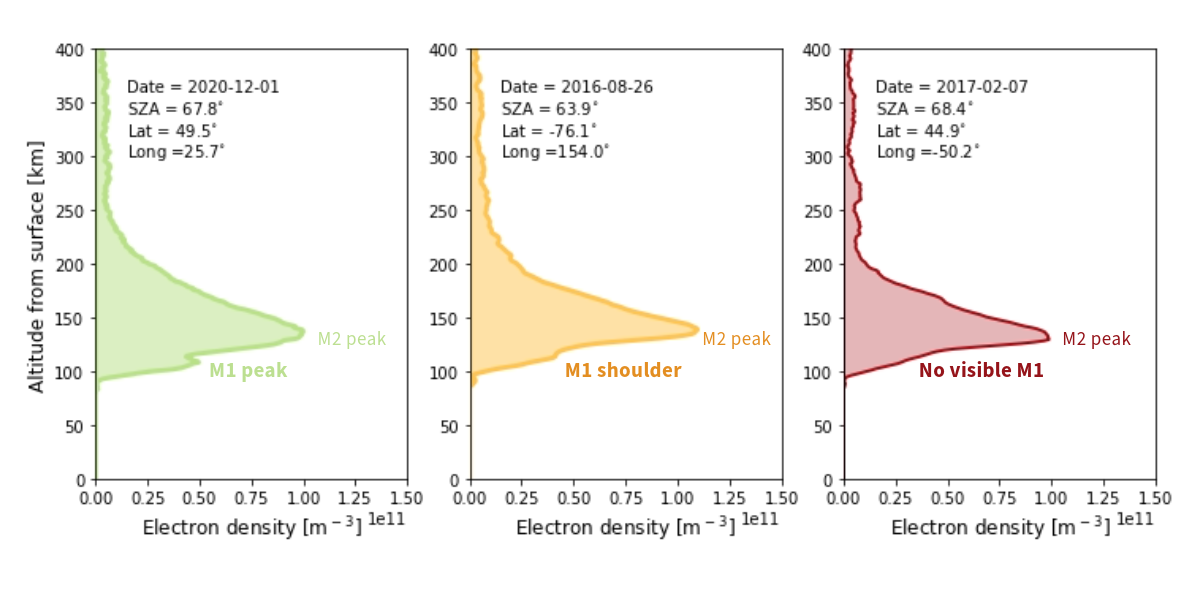}	
	\caption{Examples of ROSE electron density profiles for the \emph{strong}, \emph{medium}, and \emph{none} categories. Left is the \emph{strong} profile (green), with a well-defined peak below the M2 layer. Middle is the \emph{medium} profile (yellow), in which the M1 shows more as a shoulder rather than a peak. Right is the \emph{none} profile (brick red), with no discernible M1 peak or shoulder.} 
	\label{fig_1}%
\end{figure}

\added{Third, because the purpose of this work is to characterize the variability of both the M2 and M1 layers, the profiles which did not display an M1 layer were set aside.} \replaced{Therefore, we}{We} categorized "complete" profiles \deleted{collected at SZA $<$ 110$^{\circ}$} into three categories: \emph{strong} (example in Figure \ref{fig_1}, green), \emph{medium} (example in Figure \ref{fig_1}, yellow), and \emph{none} (example in Figure \ref{fig_1}, brick red).  
The profiles that display an obvious, well-defined M1 layer fall into the \emph{strong} category. In the \emph{medium} category belong profiles that display an M1 layer as a shoulder rather than a distinct peak, or where there was some ambiguity in distinguishing a local point of maximum density. In these cases the location of the M1 peak is less certain. In the \emph{none} category fall all the profiles that show no visible M1 peak.  \deleted{Because the purpose of this work is to characterize the variability of the M2 and M1 layer,} \replaced{O}{o}nly electron density profiles in the \emph{strong} and \emph{medium} categories were further examined.

% this confuses me a bit - we didnt explicitly filter out SZA > 110, but it was a resut of there not being any profiles with an M1 layer beyond there
\added{An interesting result of this categorization process is that all profiles used (i.e. those in the \emph{strong} and \emph{medium} categories) have SZA $<$ 110$^{\circ}$. This means that there were no profiles in this dataset with both M1 and M2 layers at SZA $>$ 110$^{\circ}$. Ionospheric structure seems to transition from dayside structure with M1 and M2 layers to nightside structure around this SZA. This also gives us an idea of }\deleted{We excluded from this study profiles collected at SZA $>$ 110$^{\circ}$: the intention behind this choice, that is including profiles collected way beyond the terminator line (SZA $=$ 90$^{\circ}$) to see} how far \replaced{past}{pass} the terminator (SZA $=$ 90$^{\circ}$) the dayside extends \added{\citep[e.g.,][]{withers2012b, Nemec:2010}}, \replaced{and if that transition occurs at different SZA at dusk compared to dawn}{and if that is different between dawn and dusk}. In other words, we would expect the dayside ionosphere at dusk to extend at SZA $>$ 90$^{\circ}$ because the ionosphere at dusk has been exposed to sunlight for longer \replaced{than}{that} it was at dawn, and, therefore,  it would take more time without direct sunlight for it to \replaced{estinguish}{disappear}. \added{From Figures 5 and 10 from \cite{Felici:2022aa} we can appreciate how, at the altitudes examined in this study, the electron density drops from $\sim O(10^{11})$ to $\sim O(10^{10})$ within $\pm 2^{\circ}$ SZA at all altitudes, once we pass the terminator.}

\begin{figure}
	\centering 
	\includegraphics[width=0.7\textwidth]{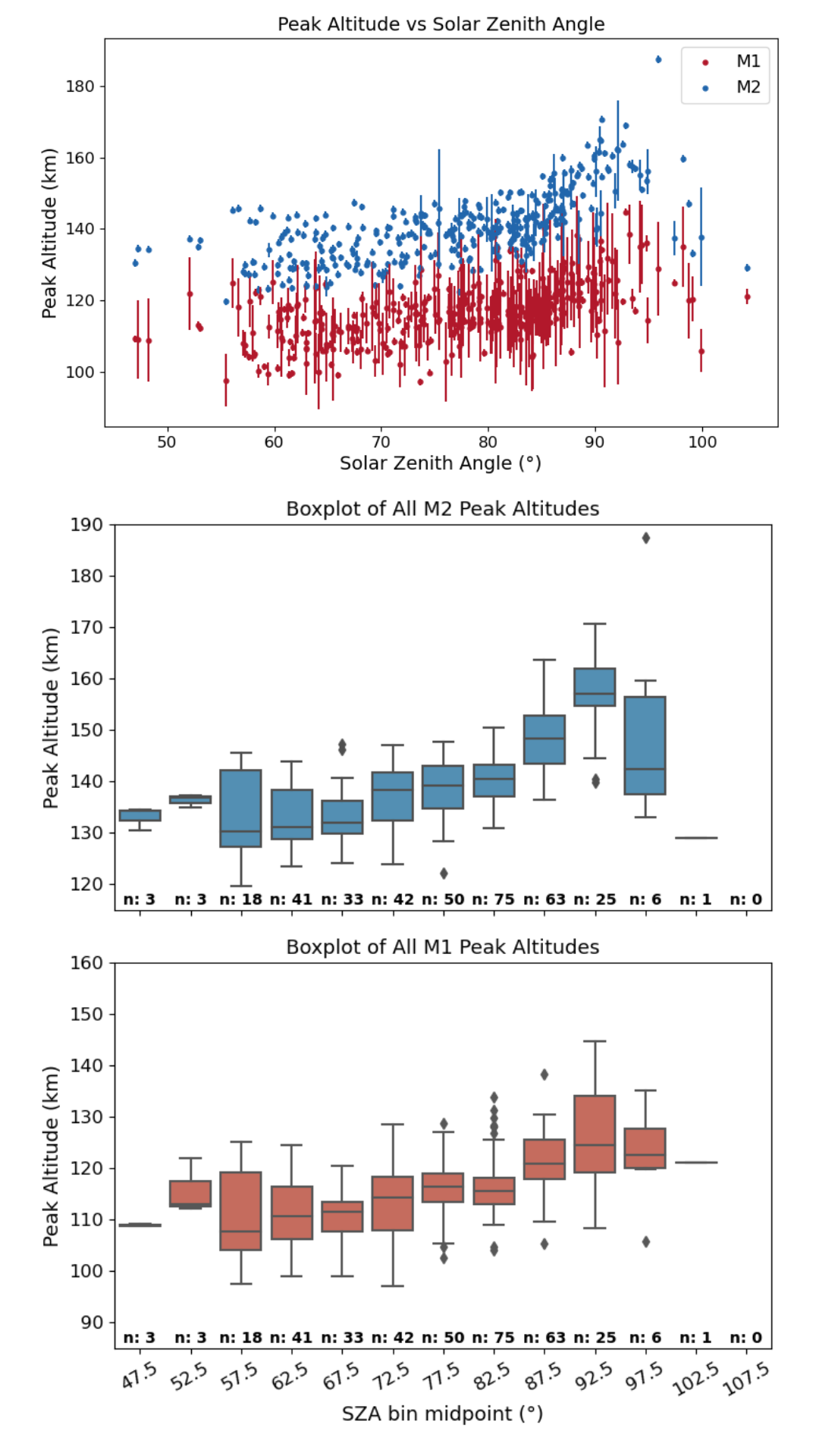}	
	\caption{ Top: M1 (red) and M2 (blue) peak altitudes of 360 electron density profiles \added{selected using criteria one to three previously listed} as a function of SZA; middle: boxplot of the 360 \replaced{M2}{M1} peak altitudes in 5$^{\circ}$ SZA bins\added{. We indicated number of profiles that fall in each bin at the bottom of the plot}; bottom: boxplot of the \replaced{M1}{M2} peak altitudes in 5$^{\circ}$ SZA bins, with center value of the bin reported on the x axis.} 
	\label{fig_2}%
\end{figure}

\begin{figure}
	\centering 
	\includegraphics[width=0.7\textwidth]{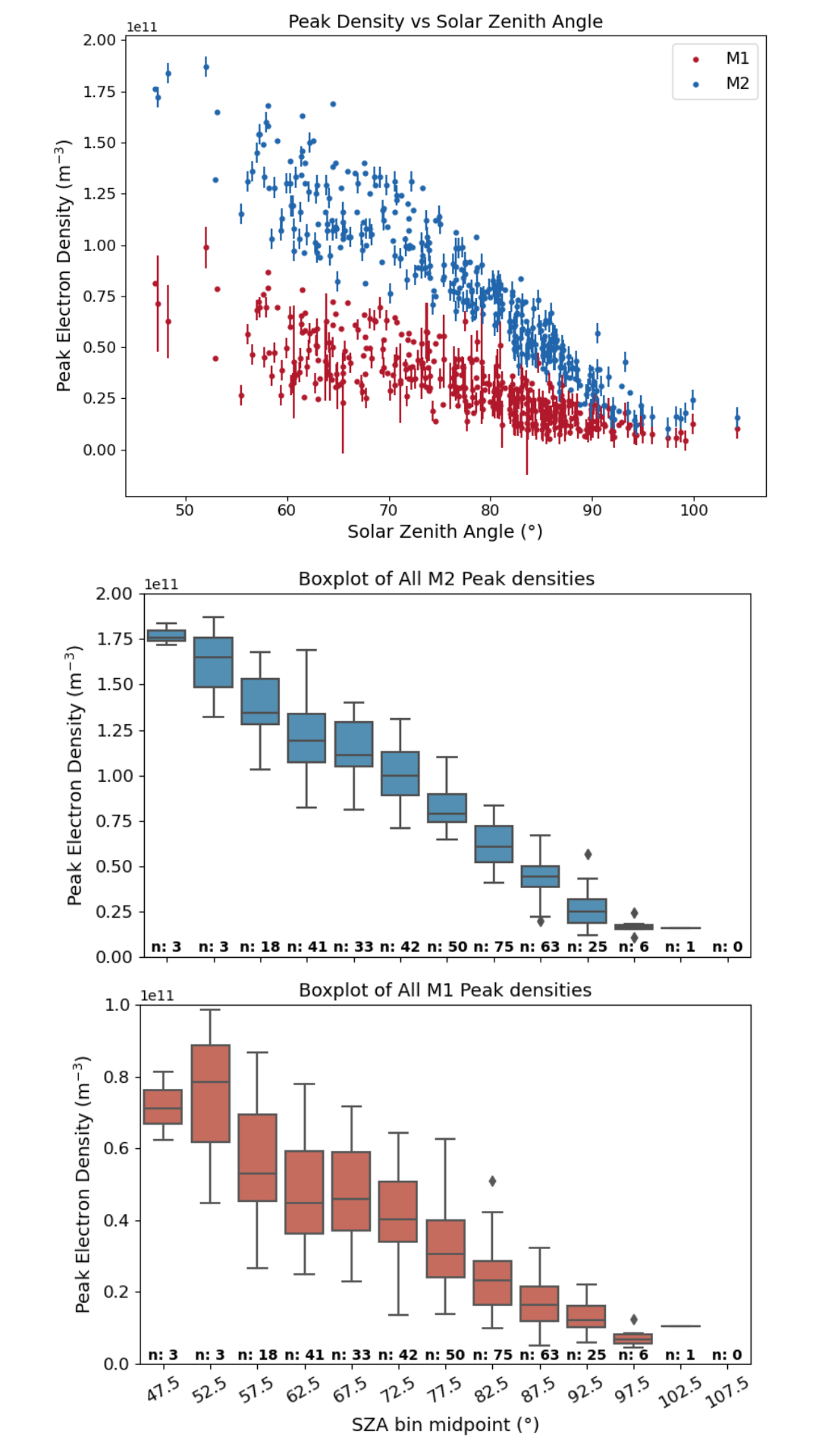}	
	\caption{Top: M1 (red) and M2 (blue) peak densities of 360 electron density profiles as a function of SZA; middle: boxplot of the 360 \replaced{M2}{M1} peak densities in 5$^{\circ}$ SZA bins\added{. We indicated number of profiles that fall in each bin at the bottom of the plot}; bottom: boxplot of the \replaced{M1}{M2} peak densities in 5$^{\circ}$ SZA bins.} 
	\label{fig_3}%
\end{figure}

\replaced{The first three criteria alone significantly decreased the number of profiles suitable for this study from 1228 to 360.}{This first level of filtering reduced the number of profiles to 360.} For these, we recorded the peak altitudes and densities of both the M1 and M2 layers with errors. \replaced{To get these values, we identified the peaks visually by inspecting each profile.}{To record peak altitude and densities of the M2 and M1 layer, we displayed one electron density profile at the time in a Graphical User Interface (GUI): through the GUI, we were able to enlarge areas of the profile, display it in linear or logarithmic scale, allowing us visually establish which point corresponded to the density peaks of the M1 or M2 layers. The GUI was built such that the clicking of the cursor on the peak of choice would automatically display the values for densities and altitudes on the terminal window with as many significant figures as the data.}  As previously mentioned, the \emph{medium} profiles do not display a clear peak, but a shoulder \added{or a cluster of smaller peaks}\deleted{, which can be smooth or present more than one local maximum}. Therefore, errors on the M1 peak in the \emph{medium} profiles were measured as the distance between two near local \replaced{maxima}{maximi} or as the width of a shoulder.  The errors on the \emph{strong} profiles were recorded as the instrument error for the density, namely 5 $\times$ 10$^{9}$ m$^{-3}$,  and as the vertical resolution of 1 km in altitude \citep{Withers:2020um}. In Figure \ref{fig_2}, top, we report peak altitude as a function of SZA for both the M1 and M2 layers in these 360 observations. In order to clearly  visualise the distribution of this dataset\deleted{, showing the median, quartiles, and potential outliers,} and compare the distribution of the M1 layer  (Figure \ref{fig_2}, \replaced{bottom}{middle}) to the one of the M2 layer  (Figure \ref{fig_2}, \replaced{middle}{bottom}), we utilize box plots\replaced{, showing the minimum (bottom bar), first quartile (Q1, bottom of the box), median (horizontal line), third quartile (Q3, top of the box), and maximum (top bar) of each 5$^{\circ}$-wide bin of the data.  We indicate the interquartile range as IQR. Outliers are indicated with diamond shapes above or below a box and correspond to data outside the maximum and minimum values interval, which is defined as the range between Q1 minus 1.5IQR and Q3 plus 1.5IQR [(Q1$-$1.5 IQR), (Q3$+$1.5 IQR).}{. A box plot is a descriptive statistics method to show minimum, first quartile, median, third quartile, and maximum of a set of data.} \added{We chose a bin width of 5$^{\circ}$ SZA to strike a balance between SZA resolution, showing the spread in the data, and reducing scatter to find if a trend existed in the median. While some of the bin populations are small, especially at the lowest and highest SZAs, we felt that these data were still valuable because they provide a sense of the general trend, particularly for low SZAs where there has not been a lot of prior study.} \deleted{The box encloses the interquartile range (IQR), namely going from the first quartile (Q1) to the third (Q3). The median is shown by an horizontal line going through the plot, and maximum and minimum are indicated as bars. Outliers are indicated with dots and correspond to data outside the maximum and minimum values interval, which is defined as the range between Q1 minus 1.5IQR and Q3 plus 1.5IQR [(Q1$-$1.5 IQR), (Q3$+$1.5 IQR)].}
\replaced{Five}{Three} features capture our attention in Figure \ref{fig_2}: a) the \replaced{IQRs}{boxes} of the M1 (red) \replaced{tend to be as large on average as}{are taller than}  the  \replaced{IQRs}{boxes} of the M2 (blue), which means a \replaced{similar}{larger} spread in altitude \added{distribution} for the values of the M1 peak compared to the M2 \added{, when the M1 is present}; b) the median of the boxes for both the M1 and M2 follow the expected trend for an idealised photochemical theory, \replaced{except for}{but} the bin centered on 52.5$^{\circ}$, which includes data collected when both solar events and dust storms were disturbing the ionosphere (see Felici et al., COMPANION MANUSCRIPT for more details); c) the bin centered on 57.5$^{\circ}$ presents a much larger box than its neighbors' boxes, for similar reasons as those listed in point two. \added{d) the drop in the median value in the bin centered on 97.5$^{\circ}$ signals the transition to the nightside ionosphere. e) in general, there are more outliers in the M1 values than in the M2 values, however we consider this number of outliers ($\sim$ 3\% of the 360 profiles) normal and not a cause for concern.} We show in Figure \ref{fig_3} the same plot as Figure \ref{fig_2}, but for peak densities of the M2 and M1 layers as a function of SZA. \deleted{What we notice here, instead, }\replaced{Besides}{besides} similar considerations to the ones we made for feature\added{s} \replaced{a) to d)}{one and three} in Figure \ref{fig_2}, it seems that peak densities for both the M1 and M2 present larger spread at low SZA, and smaller \added{spread} at higher SZA, when they inevitably start converging to \replaced{a lower order of magnitude, consistent with what was found in previous studies \citep[e.g.,][]{Felici:2022aa}}{$\sim 0$}.

\added{The fourth, and last, criterion to filter the data examined in this study was excluding from this study ROSE data collected during known dust storms and known solar events. We want to remind the reader that the} objective of this study is to characterize how the undisturbed Martian ionosphere changes with SZA, but also with L$_{S}$ and LST, in order to obtain an accurate baseline for the ionosphere \replaced{around}{at} solar minimum. This is paramount to fully understand the behavior of the Martian ionosphere, and be able to accurately quantify the effects that events external to the ionosphere (e.g. ICMEs, dust storms) have on the ionosphere. To achieve our objective, firstly, we excluded from this study profiles collected during known dust storms in MY 33, 34, 35, and 36, as those are known to heavily affect peak altitudes \cite[][and references therein]{felici2020}; secondly, we excluded profiles collected during known solar events; solar events can affect electron densities in the ionosphere, however, their effect on the lower ionosphere has not been fully characterized [see Felici et al., COMPANION MANUSCRIPT]. To appreciate the spread that these kinds of events can introduce in the data, please notice the medians for both the M1 and M2 peak altitudes (Figure \ref{fig_2}) and densities (Figure \ref{fig_3}) and their deviation from the trends expected from an idealized photochemical theory, as previously discussed. %After examination of the specific profiles, dust maps, and solar wind conditions, we concluded that this particular spread was caused by an ongoing dust storm coupled with solar events, and the ionospheric response was characterized in the COMPANION MANUSCRIPT [see Felici et al, COMPANION MANUSCRIPT]. 

\begin{figure}
	\centering 
	\includegraphics[width=1.\textwidth]{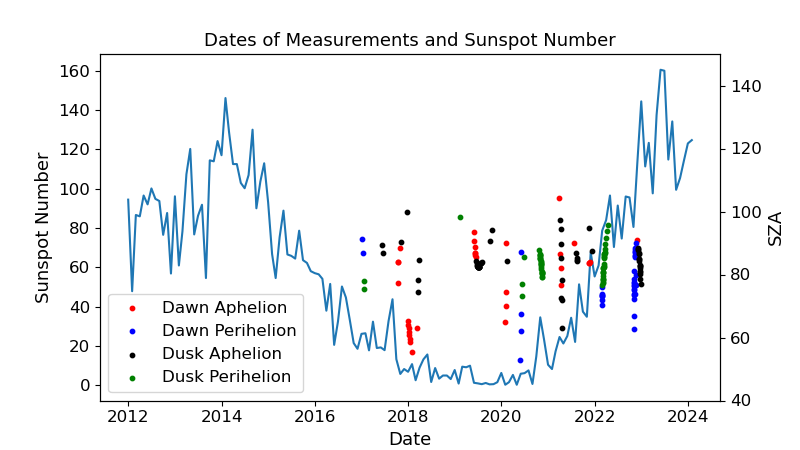}	
	\caption{Distribution of \replaced{the 219 ROSE electron density profiles which satisfied our criteria, divided into four categories }{the four categories of ROSE data} \added{-dawn aphelion, dawn perihelion, dusk aphelion, and dusk perihelion - } as a function of time and relative Sun Spot Number (SSN) and SZA. \added{The y axis for the ROSE data is on the right, labelled SZA, and the y axis for the sunspot number is on the left. The source for the sunspot number is WDC-SILSO, Royal Observatory of Belgium, Brussels.}} 
	\label{fig_4}%
\end{figure}

\added{The number of ROSE electron density profiles which satisfied all four of these criteria is} 219\deleted{ROSE electron density profiles passed this second level of filtering}, with which we will show how the ratio between M1 and M2 peak densities and the difference between the M2 and M1 peak altitudes change with SZA (see Section \ref{sec:results}). %After this second level of filtering, the total number of electron density profiles is 219, out of the 1228 with which we started. 
%To further investigate  possible causes of the spread in the peak densities and altitudes reported in previous studies, and also seen in  Figure \ref{fig_2} and \ref{fig_3}, 
\replaced{Additionally, we separated}{To achieve our objective, however, we needed to further separate} these 219 profiles into \added{four} groups based on their Local Time (LST) and Solar Longitude (L$_{S}$). %, being mindful to not define too many groups that too small a dataset would be left in each group.    
\deleted{In fact, we expect the ionosphere to be affected by these factors.} At dusk, the Martian atmosphere has been exposed to the Sun for several hours, \replaced{while for the dawn atmosphere}{differently from the dawn atmosphere, for which} the Sun has just risen\replaced{, and}{. Additionally,} depending on where Mars is in its orbit,  the planet will be exposed to different solar flux and heat \replaced{when}{depending on whether} it is at its farthest point from the Sun (aphelion)\replaced{than when it is at its}{, or at the} closest point to the Sun \deleted{along its orbit }(perihelion). %With these changing distances, the temperature of Mars’ atmosphere also changes, which could also affect ionospheric structure. 
In this study, we defined dawn as the LST between 0 and 12, and dusk as the LST between 12 and 24. We defined L$_{S}$ from 341$^{\circ}$ to 360$^{\circ}$ and 0$^{\circ}$ to 161$^{\circ}$  as "near aphelion", and from 161$^{\circ}$ to 341$^{\circ}$ as "near perihelion".   \replaced{For simplicity, in the rest of the paper, we will refer to the four groups in which the 219 ROSE electron density profiles were split as}{Therefore, we split our remnant  219 profiles  in four groups:} dawn aphelion, dawn perihelion, dusk aphelion, and dusk perihelion. Each group covers a different range of SZAs. We summarize the nomenclature, the LT, L$_{S}$, and SZA ranges of these four groups in Table \ref{table:1}. In Figure \ref{fig_4} we report the distribution of the profiles in each of these groups in LT sector, L$_{S}$ season, SZA, and phase in the solar cycle to show the relatively similar dawn-dusk coverage, albeit with some differences in SZA sampling for the same time of the year.

\begin{table}
    \centering
    \caption{Ranges in LST\replaced{ and}{,} L$_{S}$\deleted{, and SZA} \replaced{defining}{covered by} the four groups in which we split the data, and name with which we will refer to these categories in this paper.}
    \label{table:1}
    \begin{tabular}{|c|c|c|}
    \hline
    Name in this paper & Local Time (LST)& L$_{S}$ [$^{\circ}$]   \\
    \hline
    dawn - aphelion & 00:00 - 12:00 & 0-161 and 341-360  \\
   dawn - perihelion & 00:00 - 12:00 & 161 to 341 \\
    dusk - aphelion & 12:00 - 24:00 & 0-161 and 341-360  \\
     dusk - perihelion & 12:00 - 24:00 & 161 to 341 \\
    \hline
    \end{tabular}
    \end{table}

\replaced{For each of the four categories, the M1 and M2 peak densities and peak altitudes were fitted as a function of SZA with functions describing the behaviour of an ideal photo-chemically produced ionosphere \citep{chapman1931a, chapman1931b}, similarly to the methodology that \cite{fallows2015a} utilised with the MGS dataset. The fit parameters so obtained were compared for the four categories (dawn aphelion, dawn perihelion, dusk aphelion, and dusk perihelion) to quantify the effects of LST and L$_{S}$ on the Martian photo-produced dayside ionosphere.}{ To quantify the effects of LST and L$_{S}$ on the Martian ionosphere, for each of these four groups we performed  fits for M1 and M2 peak densities and peak altitudes as a function of SZA, using a methodology similar to that followed by Fallows et al., 2015 with the MGS dataset.} However, the MGS dataset had a \replaced{much}{way} narrower SZA coverage ($\simeq70^{\circ}$ to  $\simeq90^{\circ}$) than the one of ROSE, and only limited to northern latitudes, unlike ROSE.
Therefore, for the peak altitude, from \cite{fallows2015a},  to fit data \replaced{with SZA $\le$}{between SZA 0$^{\circ}$ and} 90$^{\circ}$ and extrapolate it to higher \added{and lower} SZA, we utilized the function 
\begin{equation}
z_{m} = z_{0} + L \times ln(Ch(\chi))
\end{equation}
where $m$ stands for either the M1 or the M2 peak, $z_{m}$ is the altitude of the peak, $z_{0}$ is the fitted sub-solar altitude of the peak, $\chi$ is the SZA, and $L$ is the fitted lengthscale which should coincide with the scale height of the neutral atmosphere, where an idealized photochemical theory applies. $Ch(\chi)$ is a correction factor \citep{chapman1931b,Smith-III:1972aa,fallows2015a} that \added{corresponds to the altitude predicted by the ideal Chapman theory at a SZA of $\chi^{\circ}$. It} reduces to $sec(\chi)$ at small SZA, returning the equation to the more familiar format reported by \cite{Hantsch:1990}.

For fitting the peak density, we used
\begin{equation}
N_{m} = N_{0}(\frac{1}{Ch(\chi)})^{k}
\end{equation}
where $m$ stands for either the M1 or the M2 peak, $N_{m}$ is the density of the peak, $N_{0}$ is the fitted sub-solar density of the peak, $\chi$ is the SZA, and $k$ is the fitted exponent. \added{Here $Ch(\chi)$ corresponds to the Chapman density predicted at $\chi^{\circ}$ SZA.}

\added{Only datapoints with SZA 90$^{\circ}$ or below were used in the fitting calculations, as ionospheric structure begins to change as SZA increases past this point. We found that fitting with datapoints at greater SZA caused the models to deviate from the data at SZA just less than 90$^{\circ}$. Excluding these points at SZA $>$ 90$^{\circ}$ from the fitting calculations allowed us to find the models that best fit the majority of the data.}

We report our results in the next Section (Section \ref{sec:results}).

\section{Results and Discussion}
\label{sec:results}

\added{After filtering, the 219 remaining profiles have slightly different coverage of SZA, Mars latitude and longitude, and local time than before. These ranges are summarized in Table \ref{table:new}.
Each category covers a broad range of SZAs, with dusk perihelion having the smallest range of about 25$^{\circ}$. Each group also covers about 150$^{\circ}$ of Mars latitude and about 330$^{\circ}$ of Mars longitude, meaning that we have data over most of the surface of Mars. Local time coverage also appears fairly broad in each group, meaning there are no clusters of data around particular times.}

\begin{table}
    \centering
    \caption{Summary of the local time, latitude, longitude, and SZA coverage in each category of the 219 profiles used.}
    \label{table:new}
    \begin{tabular}{|c|c|c|c|c|}
    \hline
    Category & Local Time (LST) & Mars Lat [$^{\circ}$] & Mars Long [$^{\circ}$] & SZA [$^{\circ}$]  \\
    \hline
    dawn - aphelion & 0.607222 - 11.9892 & -63.6 - 88.9 & -160.5 - 170.4 & 55.5 - 104.2   \\
   dawn - perihelion & 4.41278 - 11.9656 & -65.4 - 74.5 & -178.2 - 171.8 & 52.9 - 91.2   \\
    dusk - aphelion & 12.6456 - 22.6533 & -77.4 - 85.8 & -154.2 - 177.0 & 63.2 - 99.9    \\
     dusk - perihelion & 12.2978 - 21.2856 & -67.9 - 75.2 & -166.6 - 168.9 & 73.2 - 98.3    \\
    \hline
    \end{tabular}
    \end{table}

\begin{figure}
	\centering 
	\includegraphics[width=0.8\textwidth]{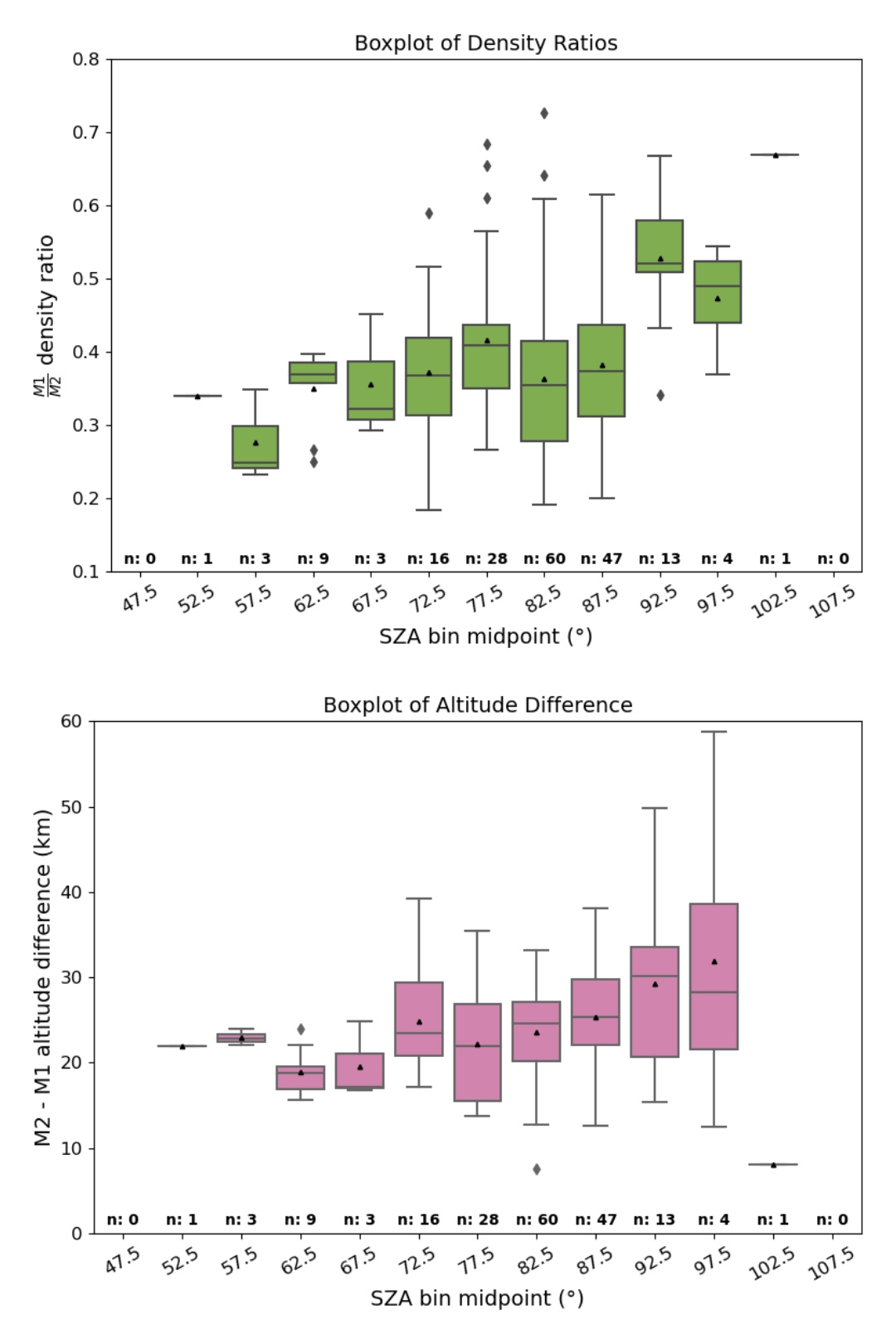}	
	\caption{At the top, box plot of the ratio between the M1 peak density and the M2 peak density.  At the bottom, box plot of the M2 and M1 peak altitude difference. \added{In both plots we indicate both mean (triangle) and median (horizontal line).}} 
	\label{fig_5}%
\end{figure}

\begin{figure}
	\centering 
	\includegraphics[width=.9\textwidth]{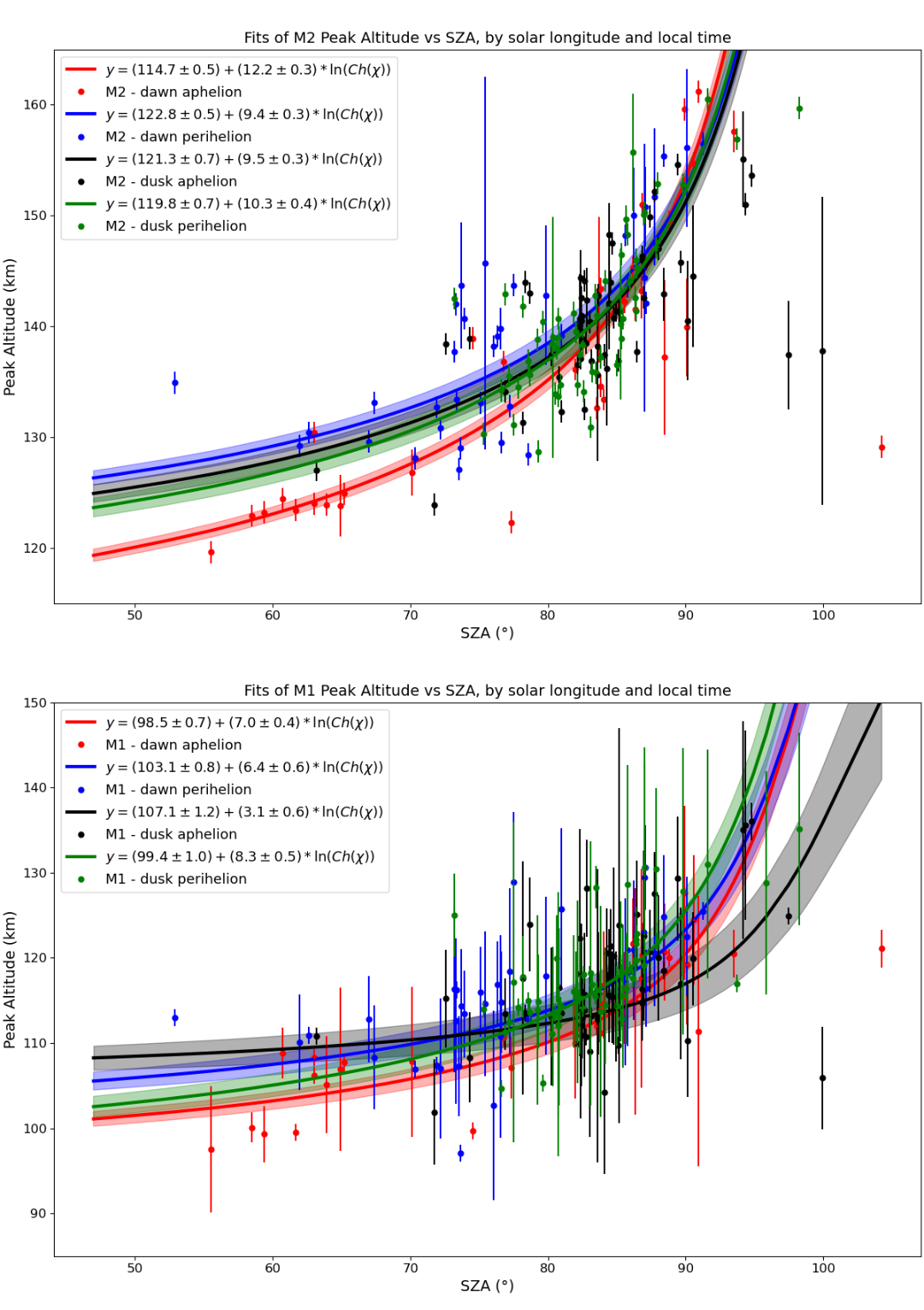}	
	\caption{Fit for M1 (bottom) and M2 (top) peak altitudes for the four groups dawn aphelion, dawn perihelion, dusk aphelion, and dusk perihelion, with the results for the fit parameters printed in the legend, and in Table \ref{table:2} for clarity. } 
	\label{fig_6}%
\end{figure}

\begin{figure} 
	\centering 
	\includegraphics[width=0.9\textwidth]{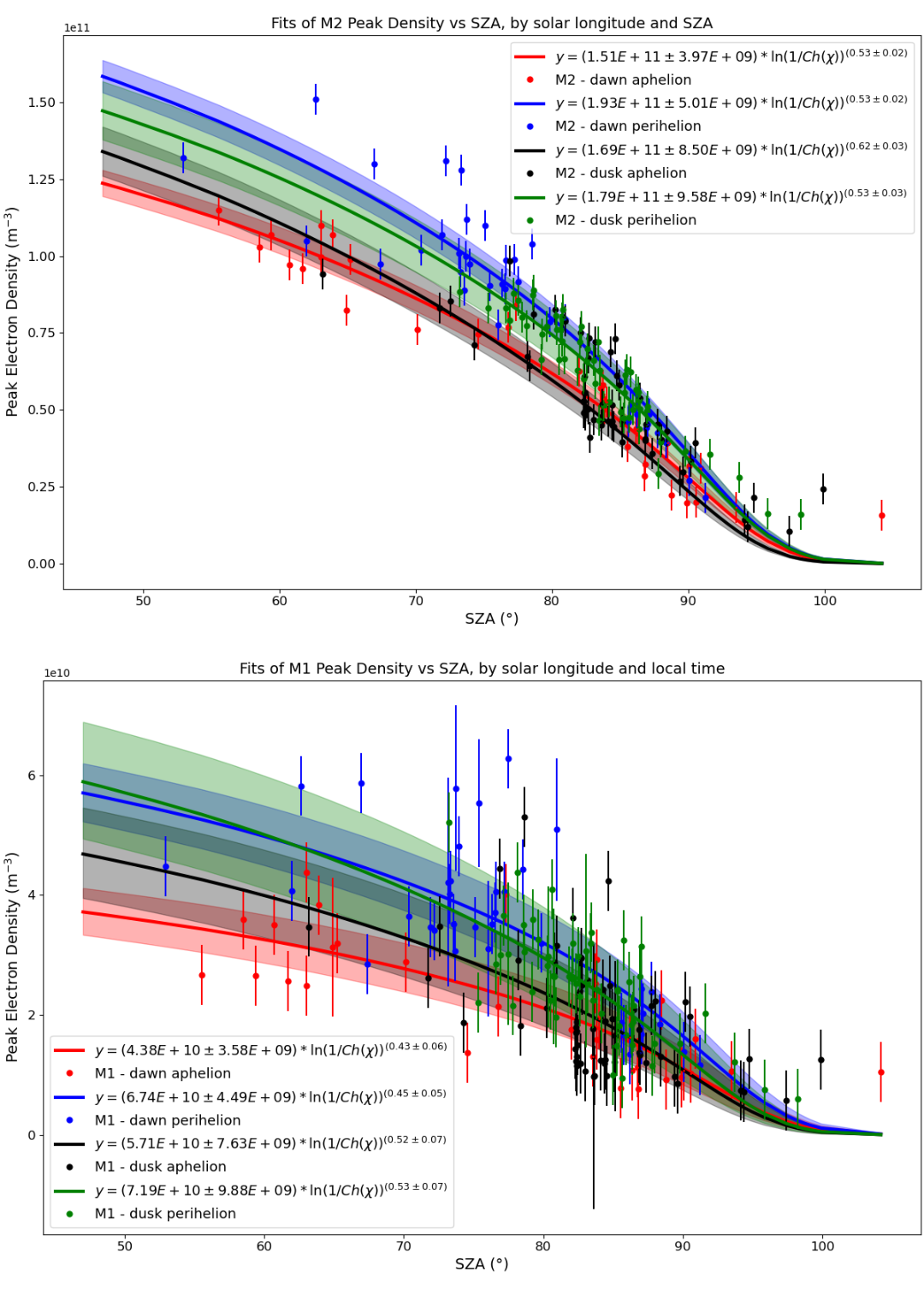}	
	\caption{Fit for M1 (bottom) and M2 (top) peak densities for the four groups\added{:} dawn aphelion, dawn perihelion, dusk aphelion, and dusk perihelion\replaced{. T}{, with t}he results for the fit parameters \added{are }printed in the legend, and in Table \ref{table:3} for clarity.  } 
	\label{fig_7}%
\end{figure}

\replaced{In Figure \ref{fig_5}, top, we report the boxplot distribution of the ratio between the M1 peak density over the M2 peak density, indicating both mean (triangle) and median (horizontal line) for a total of 219 profiles. In Figure \ref{fig_5}, bottom, we report the distribution of the difference between the M2 peak altitude and the M1 peak altitude, for a total of 219 profiles. We did not split the 219 profiles into the four groups reported in table \ref{table:1}, as we found that this made bin populations in some cases too small to be able to build a full picture for these two proxies (ratio of densities and altitude difference). Leaving these data uncategorized causes a greater spread than there would be otherwise, which introduces a few outliers in each plot, but as these outliers are only a small portion of the dataset we do not think they are a cause for concern. From \ref{fig_5}, top, we can observe that the ratio between the M1 peak densities tends to generally increase with SZA, indicating that the density of the M2 layer drops at a higher rate than the M1 density does, making the M1 more prominent at higher SZA. However, some bins at low SZA are less populated than the bins for SZA $>$ 72.5, and bins $<62.5$ are populated only by data collected in the dawn sector, therefore might not be representative of the average trend of these proxies (see \ref{fig_7} and relative text). From \ref{fig_5}, bottom,  the two layers are generally farther apart with increasing SZA, and similar considerations as the ones we did for the ratios can be made here  (see Figure \ref{fig_6} and relative text).}{In Figure \ref{fig_5}, top, we report the boxplot distribution of the ratio between the M1 peak density over the M2 peak density,  for a total of 219 profiles:  the ratio between the M1 peak densities tends to generally increase with SZA, indicating that the density of the M2 layer drops at a higher rate than the M1 density does, making the M1 more prominent at higher SZA. In Figure \ref{fig_5}, bottom, we report the distribution of the difference between the M2 peak altitude and the M1 peak altitude, for a total of 219 profiles: the two layers are generally more far apart with increasing SZA.}

%having too diverse a bin coverage between groups (see Table \ref{table:new}) to be able to build a full picture for these two proxies (ratio of densities and altitude difference) or too poor bin coverage in certain groups to actually trust those bins. 

In Figure \ref{fig_6} we show the M1 (bottom) and M2 (top) peak altitudes for the four groups dawn aphelion, dawn perihelion, dusk aphelion, and dusk perihelion, and the relative fitted curves, with fit parameters reported in the legend of Figure \ref{fig_6} and in Table \ref{table:2} in the context of other studies. 
For the M1 layer, dusk aphelion seems an outlier compared to the other trends:
in the M1 peak altitude data, the dusk aphelion fit is elevated significantly from the other categories, and has a much smaller rate of change with SZA. This may be due to a lack of data in this category between 50 - 70$^{\circ}$  SZA. %The altitude fit may be more sensitive to this absence of data than the density fit. 
\replaced{M}{Additionally, m}ost of the data for this category are concentrated between 80 - 90$^{\circ}$ SZA. This means that the data we have may not be representative of lower SZAs and the fit there may be \added{more }inaccurate. Any higher than normal data points would have a greater effect on the fit. 
Our results suggest that dawn aphelion is the group that presents consistently \added{-for both M1 and M2-} the lowest peak altitude, quite separated from dawn perihelion (more on this later). The $L$ fit parameter is consistent with a scale height \replaced{of 10 km from an}{consistent with an} idealized photochemical  theory \deleted{for the neutral atmosphere of 10 km} for the M2 layer, higher than the values obtained by \cite{fallows2015a}, and closer to those of \cite{fox2012}.
The \added{MGS} data \cite{fallows2015a} utilised was more limited in SZA (only between 70 and 90$^{\circ}$), the observations were collected at a different phase of the solar cycle (solar maximum), and effects of dust and solar events were not removed from the dataset: that is to say that we expect photo-produced ionospheric layers to behave according to an idealized photochemical theory, whereas we do not expect that the effects other phenomena have on M1 and M2 peak altitude and densities to be SZA dependent. This discrepancy in values between our $L$ fit parameters and those obtained by \cite{fallows2015a} suggests that effects on the ionosphere from solar maximum, dust storms, and solar events "mitigate" the trend the M2 and M1 peak altitude would have as a function of SZA. \added{Similar considerations can be made for the discrepancies with \cite{fox-weber2012}, who also utilised MGS data.} We also want to highlight that peak altitudes for the M1 layer can fall below 100 km, making it impossible to detect for the altitude coverage of the Viking probe, supporting the hypothesis Withers et al.,  [2023, SUBMITTED] make.

\begin{table}
    \centering
    \caption{Fit parameters for M1 and M2 peak altitudes extrapolated to the subsolar point from this study, and comparison with previous studies by other authors. We indicated with This Work with TW, for short.}
    \label{table:2}
    \begin{tabular}{|p{4cm}|p{2cm}|p{2cm}|p{2cm}|p{2cm}|}
    \hline
     Source & \small{M2 Z$_{0}$ [km]} & \small{M1 Z$_{0}$ [km]} & M2 L [km] & M1 L [km] \\
    \hline
    \footnotesize{\citet{fox2012}} & 125	& 109	& 7.2	&4.0\\
    %\footnotesize{\citet{fox-yeager2009}} & $1.9\pm0.03$ & $8.7\pm0.3$ & $0.49\pm0.01$ & $0.50\pm0.03$\\
    %\footnotesize{\citet{nemec2011}} & 1.59$\pm$0.016 & - & 0.546$\pm$0.001 & - \\
    \footnotesize{\citet{fallows2015a}} & 130.9 $\pm$ 1.8 &	104.9 $\pm$ 1.8	& 5.2 $\pm$ 1.5	& 2.5 $\pm$ 1.5\\
     \footnotesize{TW - dawn aphelion} & 114.7$\pm$0.5 & 98.5 $\pm$ 0.7& 12.2$\pm$0.3 & 7.0 $\pm$ 0.4 \\
    \footnotesize{TW - dawn perihelion} & 122.8$\pm$0.5 & 103.1$\pm$0.8 & 9.4$\pm$0.3 & 6.4$\pm$0.6 \\
    \footnotesize{TW - dusk aphelion} & 121.3$\pm$0.7 & 107.1$\pm$1.2 & 9.5$\pm$0.3 & 3.1$\pm$0.6 \\
    \footnotesize{TW - dusk perihelion} & 119.8$\pm$0.7 & 99.4$\pm$1.0 & 10.3$\pm$0.4 & 8.3$\pm$0.5 \\
    \hline
    \end{tabular}
    \end{table}

In Figure \ref{fig_7} we show the M1 (bottom) and M2 (top) peak densities for the four groups dawn aphelion, dawn perihelion, dusk aphelion, and dusk perihelion, and the relative fitted curves, with fit parameters reported in the legend of Figure \ref{fig_7} and in Table \ref{table:3} in the context of other studies. 

As for peak \replaced{densities}{altitudes}, dawn aphelion is the lowest curve, \added{with densities }well below \added{those of} dawn perihelion, while the two seasons are consistent within error for both the M1 and M2 for dusk. In other words, between aphelion and perihelion, the dawn LST sector is the one that shows the largest difference in peak density. This is consistent with what \cite{Felici:2022aa} postulated from comparing MAVEN ROSE and MAVEN LPW electron density at dawn and dusk: the two datasets were consistent within error at dusk, but not at dawn; that study utilised only data collected between 0-180$^{\circ}$ L$_{S}$, shifted from the range we considered for aphelion in this study  (see Table \ref{table:1}) of about 20$^{\circ}$, possibly capturing some of the effects that dust season itself induces on peak altitudes and densities at dawn. \added{Moreover, the higher peak densities close to perihelion are consistent with measurements of solar electron content which follows seasonal periodicity and peaks at L$_{S} \sim 240$ \citep{Sanchez-Cano:2018aa}, namely at Northern Autumn and Southern Spring.}

Peak densities and altitudes for both the M1 and M2 layers are lower than what \cite{fallows2015a} found in their study\added{ at aphelion and can be consistent within error at perihelion (with a caveat on the M1 peak altitude at dusk aphelion).} \replaced{The MGS data utilised by \cite{fallows2015a} in their studies was collected during solar maximum}{, which was conducted on data collected by MGS at solar maximum}, and at higher latitudes, suggesting that either at solar minimum electron densities are lower and the M1 and M2 peak densities occur at lower altitudes, or that latitudes (and differences in crustal \replaced{field strength and configuration}{fields}) could also affect electron densities in the lower ionosphere. \added{However, while the first statement is consistent to what \cite{Withers:2023aa} found, the second statement is not consistent with what \cite{flynn2017} found, suggesting the first explanation is the most plausible.} \added{Similar considerations can be made for the discrepancies with \cite{fox-yeager2006, fox-yeager2009}, who also utilised MGS data. As for the estimate of \cite{nemec2011}, possibly from a similar phase of the solar cycle, their estimate was conducted over more than 30000 Mars Express - MARSIS profiles, therefore averaging several effects.}

Finally, extending beyond SZA $\sim90^{\circ}$, we see mostly data collected at dusk, both perihelion and aphelion, suggesting that our hypothesis might be correct: the dusk ionosphere tends to persist longer \replaced{beyond}{passed} the terminator \added{than the dawn does}. We want to be cautious, however, given that there might be a sampling bias, and more data needs to be collected to make definitive statements.

%\section{Discussion}
%\label{sec:disc}
% table code
% 5 wide 10 tall

 %The increased amounts of EUV photons and soft X-rays ionize more neutral atoms, causing an increase in electron density. Generally, when the values of peak density are greater, the altitude of the peak is greater as well.

%dawn perihelion is also slightly higher in density and altitude than dusk. dawn aphelion densities and altitudes, however, are lower than the other three categories. Other studies of ionospheric layers have found that the dawn tends to be more variable than the dusk [CITE] (Pilinski et al., 2019). This variation is evident in the data as in the dawn perihelion and aphelion categories there are points significantly above and below the fit line for both the M1 and M2 layers, particularly at lower SZA.
%The trend that dawn aphelion and perihelion have a greater difference in altitude and density than the dusk categories could mean that the atmosphere is more sensitive to solar flux in the dawn.
%atmosphere more sensitive to solar flux in the dawn?

%NEED TO FIX M2 NUMBERS !!!
\begin{table}
    \centering
    \caption{Fit parameters for M1 and M2 peak electron densities extrapolated to the subsolar point from this study, and comparison with previous studies by other authors. We indicated \deleted{with} This Work with TW, for short.}
    \label {table:3}
    \begin{tabular}{|p{4cm}|p{2cm}|p{2cm}|p{2cm}|p{2cm}|}
    \hline
    Source & \small{M2 N$_{0}$ [$10^{11}$ m$^{-3}$]} & \small{M1 N$_{0}$ [$10^{10}$ m$^{-3}$]} & M2 k & M1 k \\
    \hline
    \footnotesize{\citet{fox-yeager2006}} & $1.82\pm0.03$ & $9.4\pm0.4$ & $0.46\pm0.01$ & $0.55\pm0.02$ \\
    \footnotesize{\citet{fox-yeager2009}} & $1.9\pm0.03$ & $8.7\pm0.3$ & $0.49\pm0.01$ & $0.50\pm0.03$\\
    \footnotesize{\citet{nemec2011}} & 1.59$\pm$0.016 & - & 0.546$\pm$0.001 & - \\
    \footnotesize{\citet{fallows2015a}} & $1.97\pm0.07$ & $7.7\pm0.7$ & $0.54\pm0.04$ & $0.55\pm0.08$ \\
     \footnotesize{TW - dawn aphelion} & $1.51\pm0.04$ & $4.38\pm0.36$ & $0.53\pm0.02$ & $0.43\pm0.06$ \\
     \footnotesize{TW - dawn perihelion} & $1.93\pm0.05$ & $6.74\pm0.45$ & $0.53\pm0.02$ & $0.45\pm0.05$ \\
        \footnotesize{TW - dusk aphelion} & $1.69\pm0.09$ & $5.71\pm0.76$ & $0.62\pm0.03$ & $0.52\pm0.07$ \\
         \footnotesize{TW - dusk perihelion} & $1.79\pm0.10$ & $7.19\pm0.99$ & $0.53\pm0.03$ & $0.53\pm0.07$ \\
    \hline
    \end{tabular}
    \end{table}
\section{Summary and Conclusions}
\replaced{219 electron density profiles of the Martian undisturbed dayside ionosphere collected by MAVEN ROSE between July 2016 and December 2022, through solar minimum leading to solar maximum, show clear M2 and M1 layers.}{We filtered 1228 electron density profiles of the Martian ionosphere collected by MAVEN ROSE down to 219 electron density profiles of the quiet ionosphere at solar minimum.} \replaced{We used these 219 profiles to characterize}{These 219 profiles show both M2 and M1 layers, allowing us to } how M2 and M1 peak electron densities and altitudes change with SZA, LST sector - dawn vs dusk - and season -\replaced{aphelion (Southern Autumn and Winter) vs perihelion (Southern Spring and Summer)}{aphelion vs perihelion}. Therefore, we split these 219 profiles into four groups: dawn aphelion, dawn perihelion, dusk aphelion, and dusk perihelion. \added{For the frequency with which the ionosphere is sampled with ROSE data, quite different from MGS RO \citep{bougher2004}, and for the difference in latitudinal sampling from MGS, tidal effects were not considered in this study.}

We find distinct differences between the different groups of data. The biggest difference, both in peak densities and in peak altitudes, is found between dawn perihelion and aphelion, consistently with the hypothesis of a more variable dawn ionosphere \citep{Felici:2022aa}. For both the M1 and M2 layers, dawn perihelion is significantly higher in altitude and greater in density than dawn aphelion. For dusk aphelion and perihelion the difference is usually less extreme. Dusk perihelion is higher in density than dusk aphelion, but dusk aphelion is higher in altitude, particularly in the M1 layer.

Densities and altitudes of the M1 and M2 layers generally tend to be higher at perihelion, expected since when Mars is closer to the Sun, we expect higher solar flux to irradiate the atmosphere. However, densities and altitudes are lower than what \cite{fallows2015a} found, at solar maximum. Namely, corresponding to the solar minimum in the solar cycle, we find lower peak densities and lower peak altitudes for both the M1 and M2 layers \citep{Withers:2023aa}.

%The M1 peak altitude can occur at altitudes lower than 100 km at low SZA, possibly the reason why previous experiments that could cover only down to 100 km altitude did not detect it. [XXX Withers 2023 SUBMITTED]

Finally, as SZA increases the M1 and M2 peaks get farther in altitude from one another, yet closer in density.

\label{sec:conclusions}
%\appendix
%Hantsch and Bauer [1990]	—	—	0.57	—
%Zhang et al. [1990]	—	—	0.57	—
%Nielsen et al. [2006]	1.79	—	0.48	—
%Morgan et al. [2008]	1.58	—	0.5	—
%Němec et al. [2011]	1.59 ± 0.016	—	0.546 ± 0.001	—
%Liao et al. [2006]	—	8.1	—	0.48
   % \citet{fallows2015a} & $1.97 \pm 0.07$ & $7.7 \pm 0.7$ & $0.54 \pm 0.04$ & $0.55 \pm 0.08$ \\
\section{Acknowledgements}
The source for the sunspot number is WDC-SILSO, Royal Observatory of Belgium, Brussels.
The Radio Occultation Science Experiment data used in this study is publicly available on the Planetary Data system. \added{This work was supported by NASA under award number NNH1OCCO4C, LASP subcontract 9500306435.}


\bibliographystyle{elsarticle-harv} 
%\bibliography{bib-m1m2paper}
%\bibliography{}





% \end{thebibliography}
\end{document}